# Intrinsic localized modes in two atomic chain; reduction of cubic anharmonicity for gap modes

V. Hizhnyakov

Institute of Physics, University of Tartu, W. Ostwaldi Str 1, 50411, Tartu, Estonia

**Abstract**

Analytical theory of large size intrinsic localized modes (ILMs) in anharmonic two-atomic chain is presented. It is shown that ILMs with frequencies close to the borders of phonon gap are govern by quartic anharmonicity while the effect of cubic anharmonicity is essentially reduced. This reduction effect is a consequence of small amplitude of vibrations of every second atom. As a result, an ILM cannot split down from the optical phonon band, in contradiction with the commonly accepted point of view. But it can spit up from the top of the acoustic phonon band making possible the existence of the ILM with the frequencies above the top of this band. It is predicted that analogous reduction of cubic anharmonicity should exist generally for even ILMs with the middle atom being at rest; it may allow the existence of ILMs with the frequencies above the top of the phonon spectrum in different lattices with realistic atomic pair-potentials.

## Introduction

Localization of vibrational excitations in periodic systems, e.g. in perfect atomic lattices is one of the basic effects of nonlinear dynamics of condensed matter [1-6]. Corresponding localized excitations are called as intrinsic localized modes (ILMs), or discrete breathers, or vibrational solitons. The idea of localization of vibrations in perfect anharmonic lattices was proposed by Kosevich and Kovalev [7] who considered a monatomic chain with nearest- neighbor harmonic, cubic, and quartic anharmonic interactions. They showed that large size localized vibrations with frequency above the top of the phonon band may exist for weak cubic anharmonicity. In [8-10] it was shown that localized vibrations of small size can also exist in anharmonic lattices. Further investigations (see, e.g. Refs. [11-17]) have revealed some basic properties of ILMs. In numerical study of ILM's, different two-body potentials such as Lennard-Jones, Born- Mayer-Coulomb, Toda, and Morse potentials have been used. All of these potentials show strong softening with increasing energy, and the ILM's found in these simulations always drop down from the optical band(s) into the phonon gap [18–21]. Exceptions may be metals [22] where due to presence of conducting electrons the interactions of atoms differ from those in insulators.

The conclusion about dropping down of ILMs from optical bands for realistic pair-potentials is essentially based on the analytical theory [7] of anharmonic modes in monatomic chain. However a priori it is not clear that in other, more complicated systems this conclusion should also hold. Here we present an analytical theory of ILMs in two-atomic chain with cubic and quartic anharmonicity. According to our theory large size ILMs cannot split up from the optical band also in a two-atomic chain with realistic anharmonic parameters. Moreover, they also cannot split down from this band, in contradiction with the commonly accepted point of view. But ILMs can spit up from the top of the acoustic band making possible the existence of the acoustic-like ILM. Such ILMs correspond to vibrations of only heavy atoms; light atoms are almost at rest for these vibrations. Analogously, vibrations with the frequency close to the bottom of the optical band would involve only light atoms; heavy atoms would be almost at rest. Due to absence of vibrations of every second atom, the effect of cubic anharmonicity is essentially reduced and only the quartic anharmonicity works. The latter anharmonicity is usually hard. This means that for vibrations with the frequency close to the border of the phonon gap the anharmonicity of working potential energy, in fact, is hard. This explains while ILMs cannot split down from the optical gap but can split up from the acoustic band. Based on this observation one can predict that that the reduction of cubic anharmonicity should take place for any even ILM with the middle atom being at rest; this should allow the existence of ILMs with frequencies above the top of the phonon spectrum in different lattices with realistic atomic pair-potentials.

## Equations of motion of atoms in two-atomic chain

Here we consider the motion of the large size self-localized anharmonic mode (ILM) in two-atomic chain with the nearest neighbors interactions described by the following potential energy:

$$W = \sum_l \left[ (K_2/2)\left((U_l - U_{l-1})^2 + (U_{l+1} - U_l)^2\right) + (K_3/3)\left((U_l - U_{l-1})^3 + (U_{l+1} - U_l)^3\right) \right. \\ \left. + (K_4/4)\left((U_l - U_{l-1})^4 + (U_{l+1} - U_l)^4\right) \right]. \tag{1}$$

Here $U_l$ is the longitudinal displacement of the atom number $l$, $K_2$ is elastic constant, $K_3$ and $K_4$ are cubic and quartic anharmonicity constants. The equations of motion of the even and odd atoms read (see, for comparison Eq. (45) in Ref. [7])

$$M_0 u_{2n,tt} = (u_{2n-1} + u_{2n+1} - 2u_{2n})\left(1 + \lambda(u_{2n+1} - u_{2n-1}) + \mu(u_{2n+1} - u_{2n-1})^2 + \mu(u_{2n} - u_{2n+1})(u_{2n} - u_{2n-1})\right), \tag{2}$$

$$M_1 u_{2n+1,tt} = (u_{2n} + u_{2n+2} - 2u_{2n+1})\left(1 + \lambda(u_{2n+2} - u_{2n}) + \mu(u_{2n+2} - u_{2n})^2 + \mu(u_{2n+1} - u_{2n+2})(u_{2n+1} - u_{2n})\right), \tag{3}$$

where $u_n = U_n/\sqrt{K_2}$, $\lambda = K_3/\sqrt{K_2^3}$, $\mu = K_4/K_2^2$, $u_{n;tt} \equiv \ddot{u}_n$ is the second derivative of $u_n$ with respect to $\tau = t/K_2^{1/4}$, $t$ is time, $M_0$ and $M_1$ are the masses of even and odd atoms; below we suppose that $M_0 > M_1$. In harmonic approximation ($\lambda = 0$, $\mu = 0$) solutions of Eq. (2) correspond to phonons with the displacements $u_{2n} = u_q e^{2iqn - i\omega_q t}$ and $u_{2n+1} = w_q e^{iq(2n+1) - i\omega_q t}$, where

$$w_q = 2u_q \cos q / (2 - M_1 \omega_q^2), \tag{4}$$

$$\omega_q^2 = M_0^{-1} + M_1^{-1} \pm \sqrt{\left((M_0^{-1} + M_1^{-1})\right)^2 - 4M_0^{-1} M_1^{-1} \sin^2 q}. \tag{5}$$

There are two phonon bands – optic and acoustic corresponding to signs "+" and "-", respectively. The frequencies $\omega_m = \sqrt{2(M_0^{-1} + M_1^{-1})}$, $\omega_a = \sqrt{2/M_0}$ and $\omega_o = \sqrt{2/M_1}$ are the top phonon frequencies of the optical and acoustic bands and the bottom of the phono frequency of the optical band, respectively. Near the ends of the phonon bands

$$\omega_{m;q}^2 \approx \omega_m^2 - 2q^2/M_+, \tag{6}$$

$$\omega_{a,o;Q}^2 \approx \omega_{a,o}^2 \mp 2Q^2/M_-, \tag{7}$$

where $M_{\pm} = M_0 \pm M_1$, $q \ll 1$ (Eq. (6)) and $Q = \pi/2 - q \ll 1$ (Eq. (7)). ILMs of large size may exist with the frequencies above the tops of the optical and acoustic bands and also below the bottom of the optical band. Corresponding wave packets of phonons consist of phonons with small $q$ or $Q$, respectively.

## ILM with the frequency above the phonon spectrum

First we consider an ILM with the frequency $\omega$ slightly above the top of the optical phonon band. We suppose that the ILM under consideration has smooth space envelope of the displacements with slowly changing amplitudes of vibrations of atoms with increasing (decreasing) of $n$. This allows one, instead of $u_{2n}$ and $-u_{2n+1}$ to consider $u(x)$ and $w(x)$, where $x$ stands for the continuous variable of space coordinate. Expanding $u(x)$ and $w(x)$ near the origin and taking into account the first two the derivatives with respect to $x$ we get the following equations:

$$M_0 u_{tt} + 2(u+w) + w_{xx} + 4\lambda(u+w)w_x + 2\mu(u+w)^3 = 0, \tag{8}$$

$$M_1 w_{tt} + 2(u+w) + u_{xx} - 4\lambda(u+w)u_x + 2\mu(u+w)^3 = 0 \tag{9}$$

(subscript $x$ denotes differentiation). Summing and subtracting Eqs. (8) and (9) we get the following equations:

$$\chi_{tt} + 4\eta\chi + \eta\chi_{xx} + \sigma\psi_{xx} + 4\lambda\chi(\sigma\chi_x - \eta\psi_x) + 4\mu\eta\chi^3 = 0, \tag{10}$$

$$M_0 u_{tt} - M_1 w_{tt} = \psi_{xx} - 4\lambda\chi\chi_x \tag{11}$$

where $\chi = u+w$ and $\psi = u-w$, $\eta = (M_0^{-1} + M_1^{-1})/2$, $\sigma = (M_0^{-1} - M_1^{-1})/2$. For $\omega - \omega_m \ll \omega_m$ Eqs. (8) - (11) can be satisfied if $|\chi|, |\psi| \ll 1$ and if $\chi(x,t) \approx \chi_0(x)\cos(\omega t)$, $\psi(x,t) \approx \psi(x)$, i.e. if the main part of $\chi$ describes the A-C-component, while the main part of $\psi$ describes the D-C-component of the ILM under consideration. In this case the left-hand side of Eq.(11) gets zero giving the relations

$$M_0 u_{tt} - M_1 w_{tt} \approx 0, \tag{12}$$

$$\psi_x \approx \lambda\chi_0^2. \tag{13}$$

Eq. (12) means that $M_0 u \approx M_1 w$, i.e. the vibration under consideration approximately conserves the center of gravity of the nearest pairs of atoms. This is a consequence of the conservation of the center of gravity by the long-wavelength optical phonons that make up this vibration. We apply the approximate relation $M_0 u \approx M_1 w$ to evaluate the small terms in Eq. (10) containing the second derivatives of $\chi$ and $\psi$ with respect to $x$. We get $\eta\chi_{xx} + \sigma\psi_{xx} \approx 2\chi_{xx}/M_+$. Using the relation $\cos^3(\omega t) = (3\cos(\omega t) + \cos(3\omega t))/4$ and neglecting the small term $4\lambda\sigma\chi\chi_x$ 9Which does not contribute) to A-C component of $\chi$, one gets Eq. (10) in the form

$$(\omega^2 - \omega_m^2)\chi_0 - 2\chi_{0,xx}/M_+ - \eta(3\mu - 4\lambda^2)\chi_0^3 = 0. \tag{14}$$

In harmonic approximation ($\lambda = 0$, $\mu = 0$) for $\chi_0 = \bar{\chi}_0 e^{iqx}$ Eq. (14) gives the dispersion relation (6), as it should be. Eq. (14) can be also presented in the following simple form:

$$X_{yy} - X + X^3 = 0, \tag{15}$$

where $X = \chi_0 \sqrt{(3\mu - 4\lambda^2)M_+\eta}/\sqrt{2}\varepsilon$, $y = x\varepsilon\sqrt{M_+/2}$, $\varepsilon = \sqrt{(\omega^2 - \omega_m^2)M_+/2}$. Solution of Eq. (15) reads $X = \sqrt{2}/\cosh(y)$. This solution gives the following envelopes of the A-C and D-C components of the immobile ILM:

$$\chi(x,t) \approx \sqrt{\frac{4M_0 M_1}{(3\mu - 4\lambda^2)M_+^2}} \frac{\varepsilon\cos(\omega t)}{\cosh(x\varepsilon\sqrt{M_+/2})}, \tag{16}$$

$$\psi(x) \approx \frac{8\sqrt{2}\lambda\varepsilon M_0 M_1}{(3\mu - 4\lambda^2)M_+^{5/2}} \tanh\left(x\varepsilon\sqrt{M_+/2}\right). \tag{17}$$

These equations in the limit $M_0 = M_1 = 1$ coincide with Eqs. (47) of Ref. [7]. Note that the condition of existence of large size ILMs above the top of the phonon spectrum of two-atomic chain $|\lambda| < \sqrt{3\mu^2/4}$ is the same as in monatomic chain (see Ref. [7]). For usual two-body potentials such as Lennard-Jones, Born-Mayer-Coulomb, Toda, and Morse potentials as well as their combinations the cubic anharmonicity parameter $|\lambda|$ is remarkably larger than $\sqrt{3\mu/4}$ [24]. Therefore for chains with usual atomic pair potentials the considered above ILMs should not exist.

## ILMs with the frequencies in the gap of phonon spectrum

Let us consider now the large size ILMs with the frequencies in the gap between the optical and acoustic bands. These ILMs are the wave packets of phonons with $Q = \pi/2 - q \ll 1$. We take into account that the vibrations with the frequencies exactly at the bottom of optical band and at the top of the acoustic band involve only heavy or light atoms. Correspondingly, for a long-wave vibrational wave-packet with the mean frequency slightly below the optical band the relations $|u_{2n+1}| \ll |u_{2n}|$ and $u_{2n+2} \approx -2u_{2n}$ hold. Analogously for a longwave vibrational packet with the mean frequency above the acoustic band the relations $|u_{2n}| \ll |u_{2n+1}|$ and $u_{2n+1} \approx -2u_{2n-1}$ hold. We denote $u_{2n} = u_o$ and $u_{2n+1} = w_o$.

Taking into account the sign-alternative dependence of $u_{2n}$ and $u_{2n+1}$ on $n$ it is convenient to use the following notations for displacements with odd and even $n$:

$$\bar{u}_{4n} = u_{4n}, \quad \bar{u}_{4n+2} = -u_{4n+2}, \\ \bar{w}_{4n+1} = u_{4n+1}, \quad \bar{w}_{4n+3} = -u_{4n+3}. \tag{18}$$

Using these notations we get for a lengthy weave-packets with the frequencies close to the borders of the phonon gap the following relations $u_{2n+2} - u_{2n} \approx -2\bar{u}$, $u_{2n+2} + u_{2n} \approx -2\bar{u}_x$, $u_{2n+1} - u_{2n-1} \approx 2\bar{w}$, $u_{2n+1} + u_{2n-1} \approx 2\bar{w}_x$. Inserting these relations to Eq. (2) we get the following equations:

$$M_0 \bar{u}_{a;tt} + 2\bar{u}_a - 2\bar{w}_{a;x} + 4\lambda\bar{u}_a\bar{w}_a + 2\mu u_a^3 \simeq 0, \tag{19}$$

$$M_1 \bar{w}_{o;tt} + 2\bar{w}_a + 2\bar{w}_{o;x} - 4\lambda\bar{u}_o\bar{w}_0 + 2\mu\bar{w}_o^3 \simeq 0, \tag{20}$$

where subscripts $a$ and $o$ stand for acoustic-like and optic-likevibrations. The terms $\bar{w}_a$ and $\bar{u}_o$ are small and can be found using harmonic approximation (4) with $q = \pi/2 - Q$. We get $(M_1\omega_a^2 - 2)\bar{w}_a \simeq 2Q\bar{u}_a$, and $(M_0\omega_o^2 - 2)\bar{u}_o \simeq -2Q\bar{w}_o$ giving

$$\bar{u}_o \approx -M_1\bar{w}_{o;x}/M_-, \quad \bar{w}_a \approx M_0\bar{u}_{a;x}/M_-. \tag{21}$$

Now Eqs (19) and (20) get the form

$$M_0\bar{u}_{a;tt} + 2\bar{u}_a + 2(M_0/M_-)\bar{u}_{a;xx} + 4\lambda\bar{u}_a\bar{u}_{a;x} + 2\mu\bar{u}_a^3 \simeq 0, \tag{22}$$

$$M_1\bar{w}_{o;tt} + 2\bar{w}_o - 2(M_1/M_-)\bar{w}_{o;xx} + 4\lambda\bar{w}_o\bar{w}_{o;x} + 2\mu w_o^3 \simeq 0. \tag{23}$$

In harmonic approximation ($\lambda = 0$, $\mu = 0$) Eqs. (22) and (23) give the dispersion relations (7) of acoustic and optical phonons near the borders of the phonon gap, as it should be. The cubic anharmonicity terms in Eqs. (22) and (23) are small as compared to this term in case of vibrations above the top of the optical spectrum (due to smallness of $\partial \bar{u}_o / \partial x$ and $\partial \bar{w}_a / \partial x$); see Eq. (14)). This result directly follows from the initial equation of motions (2) and (3): the $\propto \lambda$ terms depend on $u_{2n+1} - u_{2n-1}$ and on $u_{2n+2} - u_{2n}$. As it was indicated above, for vibrations at the top of the acoustic band $u_{2n+1} = 0$, i.e. light atoms are at the rest; analogously for vibrations at the bottom of the optical band $u_{2n} = 0$, i.e. heavy atoms are at the rest. As a result, in both these cases the working potential energy of moving atoms is even; its cubic anharmonicity is cancelled and the the $\propto \lambda$ terms disappear. Correspondingly for large-size ILMs consisting on vibrations with the frequencies close to the borders of the phonon gap the cubic anharmonicity is almost canceled and its effect is strongly reduced. This explains the smallness on the $\propto \lambda$ terms in Eqs. (22) and (23).

Due to smallness, the $\propto \lambda$ terms in Eqs. (22) and (23). can be taken into account by the method iterations. In the zero'th approximation one can put $\lambda = 0$. Then, with proper choice of variables these equations can be presented in the considered above form (15) giving the following solutions:

$$\bar{w}_a(x,t) \simeq \frac{2\varepsilon_a}{\sqrt{3\mu}} \frac{\cos(\omega t)}{\cosh\left(x\varepsilon_a \sqrt{M_-/2}\right)}, \tag{24}$$

$$\bar{u}_o(x,t) \simeq \frac{2\varepsilon_o}{\sqrt{-3\mu}} \frac{\cos(\omega t)}{\cosh\left(x\varepsilon_o \sqrt{M_-/2}\right)}, \tag{25}$$

where $\varepsilon_{a,o} = \sqrt{(\omega^2 - \omega_{a,o}^2) M_-/2}$. Now, using Eqs. (21) one can easily find the DC-components of the ILMs under consideration. These component depend on the small frequency difference parameters $\varepsilon_{a,o}$ in third power and, therefore are small. Usually quartic anharmonicity is hard: $\mu > 0$. Therefore one should expect that large-size ILMs in two-atomic chains with realistic pair potentials should exist above the top of the acoustic band while such ILMs below the bottom of the optical band cannot exist. These conclusions are in contradiction with the commonly accepted point of view [1-6, 9-14] according to which ILMs should split down from the bottom of the optical band.

## Summary

Here we present an analytical theory of large size ILMs in two-atomic anharmonic chains. According to our theory, for pair-potentials with realistic parameters the anharmonicity cannot cause the splitting an ILM up from the optical phonon band. Moreover, it cannot split an ILM down from this band either; but it can split the ILM up from the top of the acoustic band, in contradiction with the commonly accepted point of view. Such an ILM involve vibrations of almost only heavy atoms; light atoms are practically at rest for this mode. Analogously, vibrations with the frequency close to the bottom of the optical band would involve only light atoms; heavy atoms would remain essentially at rest. Due to absence of vibrations of every second atom, the cubic anharmonicity is almost switched-off, and only the quartic anharmonicity works (see the $\propto \lambda$ terms in Eqs. (2) and (3) describing the dependence of the anharmonic force, acting to an atom on the displacements of the nearest atoms; these displacements are small for vibrations with the frequency close to a border of phonon gap). The quartic anharmonicity is usually hard. Therefore for a vibration with the frequency close to the border of the phonon gap the working pair-potential, in fact, is hard. This explains why ILMs cannot split down from the optical gap but can split up from the acoustic band. Based on this consideration, one can predict that reduction of the effect of cubic (odd) anharmonicity may take place for any even ILM with the middle atom being at rest; this should allow the existence of ILMs with frequencies above the top of the phonon spectrum in different lattices with realistic atomic pair-potentials.

Finally we would like to stress that the softness of atomic pair-potentials does not guarantee that anharmonicity should reduce the mean frequency of atomic vibrations with increasing their amplitude. The actual effect of anharmonicity on the mean frequency depends on the origin of the vibration and may lead both, to decreasing and to increasing of it.

## Acknowledgement.

This work was supported by institutional research funding (IUT2-27) of the Estonian Ministry of Education and Research.